\documentclass{article}
\usepackage[latin1]{inputenc}
\usepackage{epsfig}
\usepackage{xspace}
\usepackage{a4wide}

\newcommand{\xl}{{\em xloops}\xspace}
\newcommand{\eps}{\varepsilon}
\newcommand{\bm}{\boldmath}
\newcommand{\titlestyle}{\em}
\newcommand{\volstyle}{\bf}

\begin{document}

\title{
  \begin{flushright}
    \normalsize MZ-TH 98-18 \\
  \end{flushright}
       Automatic Feynman diagram calculation with \xl\\
        --a short overview}

\author{L. Brücher\footnote{email:bruecher@thep.physik.uni-mainz.de}}

\date{27.3.1998}

\maketitle

\begin{abstract}
\xl is a program package that calculates Feynman diagrams by using
computer algebra systems. In this paper it is shown which problems
to be solved by computer algebra arise during such calculations, and
how this problems are handled in the framework of \xl.   
\end{abstract}

\section{Introduction}

\begin{figure}[htbp]
  \begin{center}
    \epsfig{file=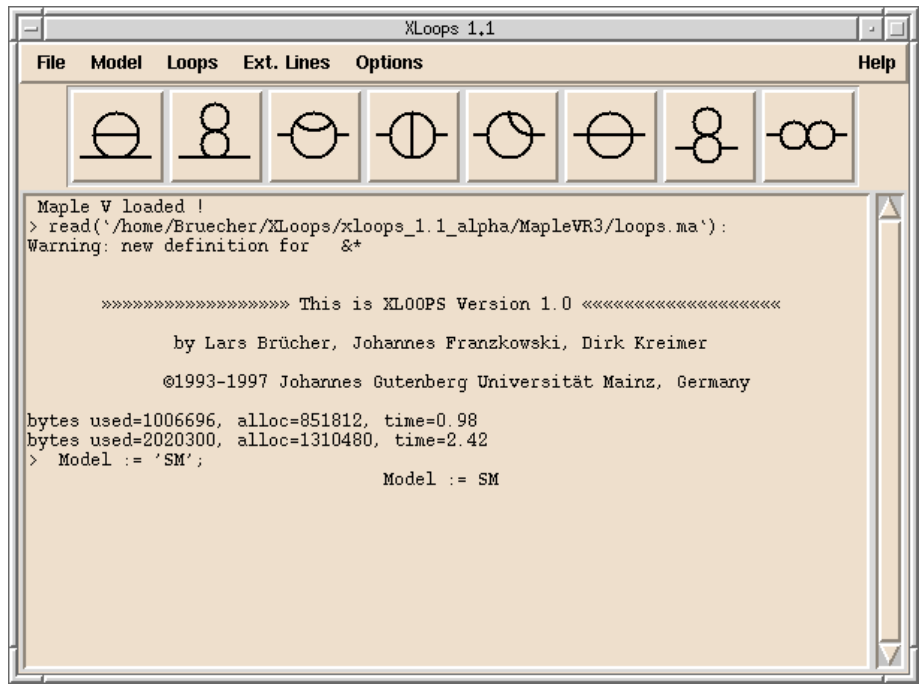,height=9cm}
    \caption{The graphical user interface}
    \label{fig:1}
  \end{center}
\end{figure}

The increasing precision of experimental data in High energy Physics
forces theorists to give very exact predictions for cross section of
reactions between elementary particles. Calculating such cross
sections efforts taking hundreds or thousands of Feynman diagrams
into account. Every such diagram represents mathematically a 4-fold
integral over internal momenta. Of course this calculations cannot be done
by hand any more. So the usage of a computer algebra system is vital.   

\section{Structure of \xl}

\begin{figure}[htbp]
  \begin{center}
    \epsfig{file=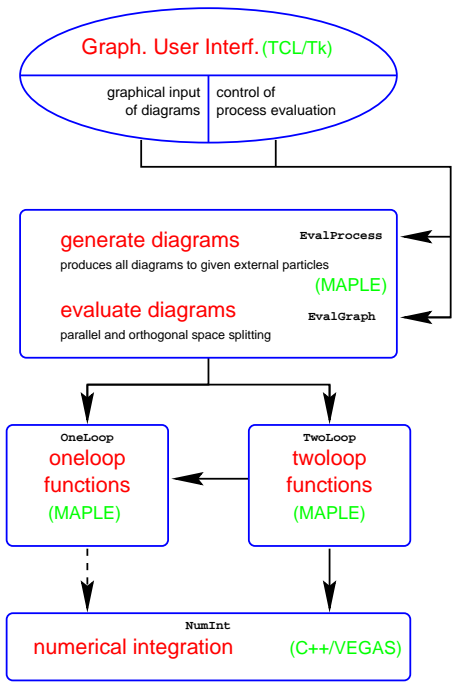,height=9cm}
    \caption{The program structure of \xl}
    \label{fig:2}
  \end{center}
\end{figure}

Building a computer program doing automatic calculations of Feynman
diagrams efforts knowledge in different sections of science. Beside
physical understanding knowledge in mathematics and computer science
is necessary to solve the arising problems. This together with just the
number of problem to be solved makes it necessary to give \xl a
structure with well defined interfaces between different programs.
How this is realized in \xl can be seen in fig. \ref{fig:2}. There can
also be seen that \xl uses different programming languages for
different tasks. First the graphical user interface (GUI) allows the
user to input the process he is interested in, as easy and comfortable
as possible. The main evaluation of the Feynman diagram is done by the
{\sf MAPLE} part of the program. Its major task is to produce
either the full result for a Feynman diagram or a result that
can be integrated numerically. Therefore \xl reduces the whole
integral to sets of standard integrals, which can be solved in an
algorithmic manner. To do so, \xl has to handle non-commutative objects like
elements of the Dirac algebra and be able to evaluate traces of such
objects. Finally these objects have to be exported to C++ code, so
that they are ready for numerical analysis.
     
\section{Physical and mathematical background}

To clarify the physical background of Feynman diagram evaluation the
following section will give a brief overview.

The quantized structure of micro cosmos is visualized by Feynman
diagram containing closed loops of internal particles. But such a
Feynman diagram is also a prescription for evaluating its contribution
to the cross section or decay width of a physical process. Every line
(propagator) and every point (vertex) corresponds to a mathematical
expression as described by the so called Feynman rules. Multiplying
these terms gives the whole contribution of the diagram. Closed loops of
particles require integration over the momenta of the particles
forming the loop. So every Feynman diagram corresponds to a $n$-fold
integral, with $n$ depending on dimension in space time and number of
closed loops. The principal analytical structure of these diagrams is:
\begin{eqnarray} \label{hier}
    \vspace{1.5cm}\mbox{\epsfig{file=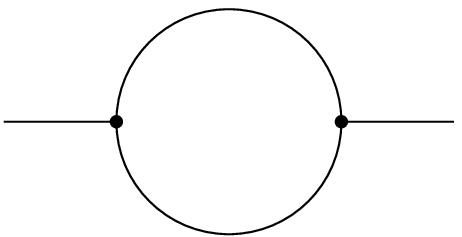,height=1cm}}
    \vspace{-1.5cm} & \longrightarrow &
    \int d^Dl \frac{l_\mu l_\nu + \slash{l} + m_1 + 
     \cdots}{[({\bm l}+{\bm q}_1)^2-m_1^2]\cdots[{\bm l}^2-m_n^2]}\ , 
\end{eqnarray}
where $\bm l$ represents a Lorenz vector and $l_µ$ represents its
components. Expanding the numerator and writing all loop-momentum
independent variables in front of each integral lead to an integral
containing just products of components of loop momentum in the
numerator and the so called Propagators ${\cal P}_i=[({\bm l}+{\bm
  q}_i)^2+m_i^2]$ in the denominator. These integrals are called
tensor integrals in contradiction to scalar integrals, which have no
loop momentum in the numerator. They can now be
further reduced to a number of standard integrals with different
techniques. 

The standard technique, called Passarino Veltman method, tries to
reduce these integrals to scalar integrals. In an intermediate step a
linear system of equations is built and has to be solved. This linear
system of equations can be rather big and so the most time consuming
part of tensor integral calculation.

An alternative technique splits the Minkowski space time into the space
spanned by the external momenta (parallel space) and its orthogonal
complement (orthogonal space). The integration over the orthogonal
space can now be transformed to an rather simple integration over the
surface of a hyper-sphere and a remaining radius integration. Beside
integrations done with residue theorem the integral formula
\begin{equation} \label{IntForm}
 \int \limits_0^{\infty} x^{\alpha-1} \, \prod_{i=1}^k (z_i+x)^{-b_i} \, dx =
  B(\beta - \alpha,\alpha) \, {\cal R}_{\alpha - \beta}({\bf b},{\bf z})
\end{equation}
expresses the result in ${\cal R}$-functions. Between these functions
exist a lot of relations, which allow to reduce all integrals to
principally one basic ${\cal R}$-function. 

A third technique uses the fact that in parallel and orthogonal space
every scalar product between an outer momentum ${\bm q}_i$ and a loop
momentum ${\bm l}$ just projects out one component of the loop
momentum. Every such product occurs in one of the propagators. So
every component of $l_µ$ from parallel space can be expressed in terms
of progators and masses $m_i$. The remaining components from orthogonal
space can be expressed in terms of the last propagator from
(\ref{hier}), which has no external momentum. So a recursive
definition, which reduces every tensor integral to its corresponding
scalar integral and number of integral with less propagators is gained.    

\section{Exemplary CA problem: reduction of tensor integrals}


In this section some techniques used by \xl should be shown in more
detail. As example serve to different methods for evaluating tensor
integrals with \xl. The standard technique (Passarino Veltman method)
is not used by \xl, as solving linear systems of equations with
computer algebra methods is much slower than the algorithms discussed
below. 
  
\subsection{A fast technique: recurrence relations with ${\cal
    R}$-functions}

In the case of the one-loop two-point function the resulting ${\cal
  R}$-functions gained by integration fulfill the following
recurrence relation:
\begin{equation}
   (b_1+b_2)\,{\cal R}_t(b_1,b_2;z_1,z_2)\ =\
        (b_1+b_2 + t)\, {\cal R}_t(b_1+1,b_2;z_1,z_2)
           \ -\ tz_1\, {\cal R}_{t-1}(b_1,b_2;z_1,z_2)
\end{equation}
resulting in the basic ${\cal R}$-function:
\begin{eqnarray}
\lefteqn{{\cal R}_{-\eps} (\mbox{$-\frac{1}{2} + \eps$},1;z_1,z_2)} \nonumber\\
& = & + \mbox{$\frac{1}{2}$} \, z_{1}^{-\eps} \, \left(\frac{z_{1}}{z_{2}}\right)
^{1-\eps} \, \left[\left(1+\sqrt{1-\frac{z_{1}}{z_{2}}}\right)^{-1+2\eps} +
\left(1-\sqrt{1-\frac{z_{1}}{z_{2}}}\right)^{-1+2\eps}\right] +
O(\eps^{2}) \nonumber  
\end{eqnarray}
This can of course be directly translated into the following {\sf
  MAPLE}-code:
{\small
\begin{verbatim}
R2 := proc(ind,b1,b2,z1,z2) local erg;
              .
              .
              .
# *** Parameter increasing ***
   elif b1<0 then
    erg := ( (2*b1+2*b2+2*ind) * R2(ind,b1+1,b2,z1,z2)
            - 2*ind*z1* R2(ind-1,b1+1,b2,z1,z2) )/(2*b1+2*b2);
              .
              .
   elif ind=0 then
    erg := z2^eps/(2*z1^(2*eps)) *
           ( (1+Sqrt(1-z1/z2))*(1-Sqrt(1-z1/z2))^(2*eps)
            +(1-Sqrt(1-z1/z2))*(1+Sqrt(1-z1/z2))^(2*eps) );
   RETURN( erg );
  fi;
end:
\end{verbatim}}
The recurrence relations are for most cases twice as fast as solving
the linear system of equations, where as intermediate step matrices
have to inverted. 

\subsection{Even faster technique: cancellation of Propagators}

Parallel and orthogonal space splitting lead for the one-loop two-point
function to the following relation (here just for parallel space component of
loop momentum):
\begin{equation}
     l_\|  =  \frac{1}{2\,q_{\|}}\left[{\cal P}_1-{\cal P}_2 + C\right]
\end{equation}
in terms of integrals this means:
\begin{equation}
  \int \frac{l_\|^n}{{\cal P}_1{\cal P}_2} =
  \frac{1}{2\,q_{\|}}\left[\, \int \frac{l_\|^{n-1}}{{\cal
  P}_2} - \int \frac{l_\|^{n-1}}{{\cal P}_1} + C \int
  \frac{l_\|^{n-1}}{{\cal P}_1{\cal P}_2}\: \right]
\end{equation}
with a integration independent constant $C$.So this equation reduces
the exponent of loop momentum of a tensor integral by 
one and produces additional, simpler integrals. Applying this formula
$n$ times ends with the scalar two-point function. This can also be
written in a iterative algorithm:   
{\small
\begin{verbatim}
FastTensor2Pt := proc(p0,p1,q0,m_1,m_2,eps) 
                 local C1,C2,DM,i0,i1,i0s,cf1,cf2,cf3,cf4,cf5,cf6,m1,m2,temp0;
                 global rho;
  if not type(p1,even) then RETURN( 0 ); fi;
  cf4 := 0;
  for i0 from 1 to p0 do
   i0s := trunc((i0-1)/2);
   cf3 := 0;
   for i1 from 0 to i0s do 
    cf1 := binomial(i0-1,2*i1)*(2*q0)^(2*i1-p0)*Two2OnePtFactor(2*i1,p1,eps);
    cf2 := (C1)^(i0-1-2*i1)*Tadpole(m2,1+i1+p1/2-eps)
           +(-1)^i0*C2^(i0-1-2*i1)*Tadpole(m1,1+i1+p1/2-eps);
    cf3 := cf3 + cf1*cf2;
   od;
   cf4 := cf4 + binomial(p0,i0)*(-C1)^(p0-i0)*cf3;
  od;
  cf6 := cf4/(1-eps);
  temp0 := (-C1/2/q0)^p0*Small_2Pt(p1,q0^2,m1,m2,eps)+cf6;
  temp0 := eval(subs(C1=q0^2-m_1+m_2,C2=q0^2-m_2+m_1,m1=m_1,m2=m_2,temp0));
  RETURN( eval(temp0) );
end:
\end{verbatim}}
Depending on the dimension of parallel space this technique is even a
bit faster than the above described. 

\section{Conclusion and outlook}

In high energy physics different techniques of symbolic calculation
are used. Some of them used by \xl have been pointed out here in
detail. Moreover \xl uses the following elements and algorithm of
computer algebra systems: 
\begin{itemize}
\item user defined recursive functions
\item power series expansion
\item fast list evaluation 
\item algebraic term collection
\end{itemize}
The hole package is still a developing project. It can be obtained
from its homepage on the WWW:
\begin{center}
{\tt http://wwwthep.physik.uni-mainz.de/$\sim$xloops}
\end{center}


\begin{thebibliography}{1}

\bibitem{Col}
{J. C. Collins}.
\newblock {\titlestyle {Renormalization}}.
\newblock Cambridge University Press (1984)

\bibitem{Car}
{B. C. Carlson}.
\newblock {\titlestyle {Special Functions of Applied Mathematics}}.
\newblock Academic Press (1977)

\bibitem{xl1}
{L. Br\"ucher, J. Franzkowski, D. Kreimer}.
\newblock {\titlestyle Introduction to \xl}.
\newblock {\titlestyle Nucl.Instrum.Meth.} {\volstyle A 389} (1997) 323--342

\bibitem{xl2}
{L. Br\"ucher, J. Franzkowski, D. Kreimer}.
\newblock {\titlestyle The \xl Manual}.
\newblock {\titlestyle Eprint} {\volstyle hep-ph/9710484}

\bibitem{xl3}
L. Br\"ucher, J. Franzkowski, D. Kreimer.
\newblock {\titlestyle xloops: Automated Feynman diagram calculation}.
\newblock {\titlestyle Comput. Phys. Commun.} {\volstyle 115} (1998) 140

\bibitem{Fr3}
{L. Br\"ucher, J. Franzkowski, D. Kreimer}.
\newblock {\titlestyle Loop Integrals, R Functions and their Analytic
  Continuation}.
\newblock {\titlestyle Mod. Phys. Lett.} {\volstyle A9} (1994) 2335--2346

\bibitem{oneloop1}
{L. Br\"ucher and J. Franzkowski}.
{\em Mod. Phys. Lett.} {\bf A14} (1999) 881;

\bibitem{oneloop2}
{L. Br\"ucher, J. Franzkowski and D. Kreimer}.
\newblock {\em Comp. Phys. Comm.} {\bf 85} (1995) 153--165;

\bibitem{oneloop3}
{L. Br\"ucher, J. Franzkowski, D. Kreimer}.
\newblock {\em Comp. Phys. Comm.} {\bf 107} (1997) 281--291.

\end{thebibliography}

\end{document}